\documentclass[conference]{IEEEtran}
\IEEEoverridecommandlockouts
\usepackage{cite}
\usepackage{amsmath,amssymb,amsfonts}
\usepackage{amsthm}
\usepackage{subcaption}
\usepackage{graphicx}
\usepackage{textcomp}
\usepackage{xcolor}
\usepackage{xspace}
\usepackage{booktabs}

\usepackage{optidef}
\usepackage{algorithm}
\usepackage{algpseudocode}
\usepackage{lipsum}
\usepackage{nomencl}
\usepackage{hyperref} 
\usepackage{color,soul}
\algrenewcommand\algorithmicindent{0.8em}

\begin{document}

\title{ISAC-over-NTN: HAPS–UAV Framework for Post-Disaster Responsive 6G Networks}
\author{\IEEEauthorblockN{Berk Ciloglu$^{1}$, Ozgun Ersoy$^{2}$, Metin Ozturk$^{2}$, Ali Gorcin$^{3,4}$}
\IEEEauthorblockA{$^1$Department of Information Engineering, University of Brescia, Brescia, Italy\\
\IEEEauthorblockA{$^2$Electrical and Electronics Engineering, Ankara Yıldırım Beyazıt University, Ankara, Türkiye}
\IEEEauthorblockA{$^3$\href{https://hisar.bilgem.tubitak.gov.tr/en/}{Communications and Signal Processing Research (HİSAR) Lab., T{\"{U}}B{\.{I}}TAK B{\.{I}}LGEM, Kocaeli, Türkiye}}
\IEEEauthorblockA{$^4$Department of Electronics and Communication Engineering, Istanbul Technical University, {\.{I}}stanbul, Türkiye}
}
}
\maketitle

\begin{abstract}
In disaster scenarios, ensuring both reliable communication and situational awareness becomes a critical challenge due to the partial or complete collapse of terrestrial networks.
This paper proposes an integrated sensing and communication~(ISAC) over non-terrestrial networks~(NTN) architecture---referred to as ISAC-over-NTN---that integrates multiple uncrewed aerial vehicles (UAVs) and a high-altitude platform station (HAPS) to maintain resilient and reliable network operations in post-disaster conditions.
We aim to achieve two main objectives: i) provide a reliable communication infrastructure, thereby ensuring the continuity of search-and-rescue activities and connecting people to their loved ones, and ii) detect users, such as those trapped under rubble or those who are mobile, using a Doppler-based mobility detection model.
We employ an innovative beamforming method that simultaneously transmits data and detects Doppler-based mobility by integrating multi-user multiple-input multiple-output~(MU-MIMO) communication and monostatic sensing within the same transmission chain.
The results show that the proposed framework maintains reliable connectivity and achieves high detection accuracy of users in critical locations, reaching 90\% motion detection sensitivity and 88\% detection accuracy.
\end{abstract}


\section{Introduction}

The integrated sensing and communication~(ISAC) framework emerges as one of the key enablers of sixth-generation of cellular networks~(6G), promising the combined use of wireless connectivity and environmental perception on a single physical layer~\cite{isac_survey_high_cited}.
It enables intelligent and environmentally aware networks with the capability of maintaining situational awareness and reliable connectivity~\cite{cop_isac}.
Especially in post-disaster situations, the ISAC paradigm can operate not only by providing connectivity but also by supporting environmental awareness in terms of locating victims and monitoring rescue operations, which are highly critical during disaster relief.
Natural disasters often severely damage or even completely disable terrestrial communications infrastructure, creating widespread communication outages and depriving affected communities and emergency response teams of reliable connectivity~\cite{ntn_disaster_magazine}.
In such situations, non-terrestrial networks~(NTN)---including uncrewed aerial vehicles~(UAVs), high altitude platforms~(HAPS), and satellites---can enable reliable connectivity and sensing services with their rapid deployment capabilities~\cite{satcom}.
UAVs can reach the required location quickly and provide reliable communications in a mobile and flexible manner.
HAPS, on the other hand, is a fast and reliable NTN node, providing ultra-wide coverage due to its unique location in the stratosphere, along with high capacity and a nearly accurate line-of-sight (LoS) characteristics.

The use of ISAC technology in conjunction with NTN provides highly efficient sensing and communication over wide areas.
ISAC, with its unique capability of combining radar and communication signals within a single framework, optimizes hardware costs and spectral efficiency.
Moreover, NTN components, particularly HAPS and UAVs are used in this paper, can both transmit data and perform environmental sensing cooperatively within the same network, thereby they offer a complementary architecture to terrestrial networks~(TN), with HAPS providing wide-area coverage, high-capacity backhaul, while UAVs provide flexible, low-altitude access links and detailed environmental sensing. 
This combination aims to enhance frequency and resource efficiency through ISAC technology, while NTN technology seeks to expand coverage and ensure connection continuity and reliability.

The studies in the literature show limited integration of ISAC-over-NTN, particularly concerning HAPS and UAVs.
In~\cite{maxminfairness}, a HAPS-assisted ISAC system is introduced, and the authors develop a non-convex optimization model for balancing beam pattern gain and signal-to-interference-plus-noise ratio~(SINR).
The HAPS-ISAC architecture is proposed in~\cite{HAPS-ISAC}, which uses advanced beamforming techniques with a multiple-input and multiple-output~(MIMO) multiple-input and single-output (MISO) configuration to provide both high accuracy in target detection and reliable connectivity for communication users. 
The study in~\cite{Optimizing} proposes an innovative ISAC system using HAPS and UAVs for 6G networks; the system reduces the computational load of UAVs by using HAPS as a central processing unit~(CPU).
In~\cite{ISAC-Assisted}, the authors propose an ISAC-assisted two-stage beam alignment method for flying ad-hoc networks (FANETs) between multiple  HAPS. 
Furthermore, in~\cite{Aerial_ISAC}, a non-terrestrial ISAC system is proposed, where a full-duplex ISAC base station~(BS) mounted on a HAPS is deployed to provide both communication and sensing services simultaneously. 
Studies on ISAC-over-NTN have focused on optimization-based designs for beamforming, power allocation, or alignment under idealized network conditions; however, these works are predominantly limited to satellite-based or UAV-only systems.
These studies generally neglect dynamic user mobility and real-time adaptation to different user situations because they make optimization-oriented designs under idealized and static NTN conditions.
Consequently, jointly implementing situationally aware ISAC operations in post-disaster environments, where communication and sensing are crucial, remains largely unexplored.

In this study, we propose a comprehensive ISAC-over-NTN framework that integrates multiple UAVs and a single HAPS node into a unified system capable of providing both communication and sensing functions in a post-disaster scenario, where terrestrial infrastructure fails.
In the proposed architecture, UAVs equipped with BSs jointly perform downlink multi-user MIMO (MU-MIMO) communication and monostatic sensing for ground users, while HAPS offers a backhaul link, extra capacity, and coverage continuity.
The proposed framework distinguishes itself from traditional approaches that rely solely on optimization by providing an end-to-end, dynamic, and sensing-aware network operation model designed to ensure communication continuity, situational awareness, and resilience under critical conditions.
The main contributions of this work are listed as follows:
\begin{itemize}
    \item In a post-disaster situation where the terrestrial BSs are destroyed, reliable communication and detection of users affected by the disaster are achieved with the ISAC-over-NTN framework, which employs MU-MIMO BSs mounted on UAVs and HAPS-based backhaul links.
    \item The UAV network architecture performs joint sensing and communication using beamforming operations under the coordination of the HAPS.
    \item Different user scenarios are considered, including victims trapped under the rubble, survivors who escaped collapsing buildings, and a group of users gathered at the emergency assembly point, to compare detection and communication performance across user types.
    \item Semi-orthogonal user selection (SUS) and a reuse-based transmission scheduler are developed to ensure spectrum-efficient and interference-aware access among UAVs.
    \item A sensing mechanism utilizing Doppler estimation and Cramér-Rao Bound (CRB) analysis is designed to identify user mobility patterns and support situational awareness and prioritized network response.
\end{itemize}


\section{System Model}\label{sec:system_model}

\subsection{Network Topology}

\begin{figure}
\centering
\includegraphics[width=.7\linewidth]{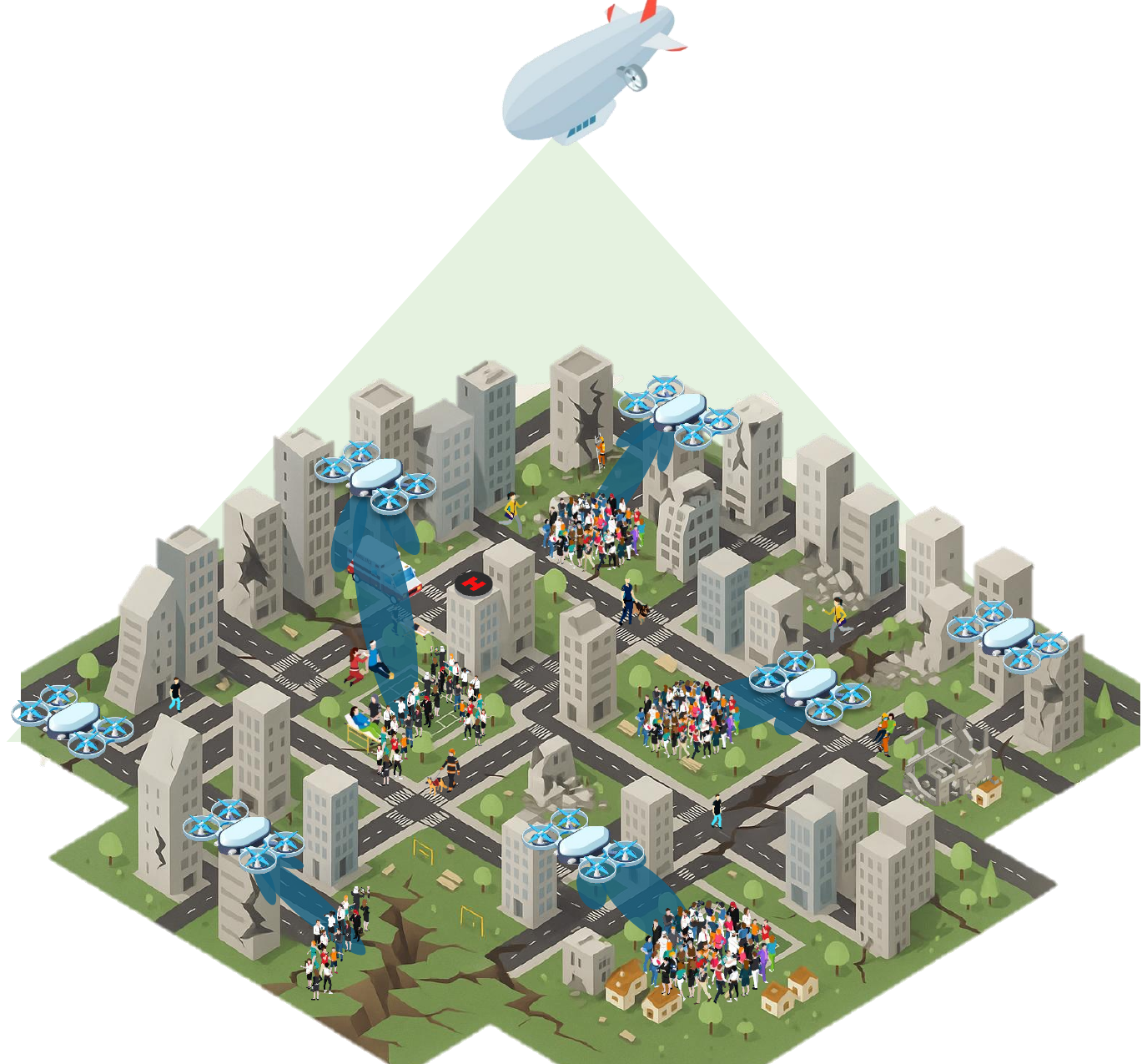}
\caption{The system model illustrates multiple UAVs performing ISAC channel under the coordination of a HAPS. 
}
\label{fig:system_model}
\end{figure}
A post-disaster scenario, in which all the terrestrial BSs have collapsed, is considered in this work. 
A multi-layer NTN architecture is developed, featuring multiple UAVs equipped with millimeter-wave (mmWave) MU-MIMO antenna-based BSs, referred to as UAV-BS, and a single HAPS serving as an International Mobile Telecommunications (IMT) BS, referred to as HIBS (the term HIBS will be adopted hereafter throughout the paper). The HIBS operates in a sub-6 GHz frequency spectrum using a single antenna.
We consider a system consisting of $U\in\mathbb{Z}^+$ UAV-BSs with an altitude of $h_u$, and indexed by $u\in\mathcal{U}$ where $\mathcal{U}=\{1, 2, ..., U\}$, a single HIBS located at the center of the considered environment at an altitude of $h_\text{h}$, where $h_\text{h} \gg h_u,~\forall u \in \mathcal{U}$.
Moreover, the system comprises $K\in\mathbb{Z}^+$ ground user equipments (UEs), indexed by $i\in\mathcal{K}$ where $\mathcal{K}=\{1, 2, ..., K\}$. 
The system model is illustrated in Fig~\ref{fig:system_model}.
The UAV-BS network supports MU-MIMO transmission, mobility-aware scheduling, and Doppler-based sensing for situational awareness during terrestrial outages.
Therefore, the main purpose of the UAV-BS network is to provide reliable communication while ensuring sensing services.
HIBS is used for extended capacity, coverage, and backhaul connectivity for the UAV-BSs.
For comparative evaluation purposes, a TN configuration with an $N_\text{b}\in\mathbb{Z}^+$ number of BSs, operating at a sub-6 GHz spectrum, is also considered to provide comparative results under different BS failure rates.

The considered duration is divided into $\Delta t$ outer slots, and each outer slot consists of several mini-slots.
A reuse scheduler manages these mini-slots, allowing groups of UAV-BSs to activate one group per mini-slot, thereby reducing co-channel interference.
The three dimensional~(3D) distance between UAV-BS $u$ and user $i$ is given by $d_{u,i} = \|\mathbf{r}_i-\mathbf{r}_u\|$ and the angular coordinates are computed as $\phi_{u,i} = \mathrm{arctan2}(y_i-y_u,\,x_i-x_u), \theta_{u,i} = \arcsin\!\left(\frac{h_u}{d_{u,i}}\right)$.
The corresponding radial velocity and Doppler frequency are expressed respectively as
$v_{r,u,i} = \dot{d}_{u,i}$, and $\mu_{u,i} = \frac{2v_{r,u,i}}{\lambda}$.
\subsection{Channel Models}
Each UAV-BS employs a uniform planar array (UPA) consisting of $M_x \times M_y$, where $M_x, M_y \in \mathbb{Z}^+$, elements with half-wavelength spacing. Thus, the total number of transmit antenna elements, $N_\text{t}\in \mathbb{Z}^+$ is calculated as $N_\text{t}=M_xM_y$. 
The transmit steering vector toward $(\theta,\phi)$ is given by~\cite{Dynamic_ISAC_UAV}
\begin{equation}
\mathbf{a}_t(\theta,\phi)
 = \frac{1}{\sqrt{N_t}}
 \exp\{-j k_0 [m d_x\sin\theta\cos\phi + n d_y\sin\theta\sin\phi]\},
\end{equation}
where $k_0$ is propagation constant, $m$ is the element index along the $x$-axis with $m=0,...,M_x-1$, and $n$ is the element index along the $y$-axis with $n=0,...,M_y-1$. 
Stacking them column-wise yields $\mathbf{a}_\text{t}(\theta,\varphi) \in \mathbb{C}^{N_\text{t}}$.
The received steering vector $\mathbf{a}_\text{r}(\theta,\varphi)$ is defined identically to $\mathbf{a}_\text{t}(\theta,\varphi)$.
The log-normal path loss model is used for the UAV-BS-UE channel, taking into account shadowing and the Rician multipath fading, as
\begin{equation}
H_{u,i} = \sqrt{\tfrac{K}{K+1}}\;\alpha_{\mathrm{LoS}}(d)\,a_t(\theta,\phi) 
+ \sqrt{\tfrac{1}{K+1}}\;\alpha_{\mathrm{LoS}}(d)\,H_{\mathrm{NLoS}},
\end{equation}
where $\alpha_{\mathrm{LoS}}$ is the path loss when the LoS region, and $H_{\mathrm{NLoS}}$ is the path loss when the non-LoS~(NLoS) region.
For the HIBS-UE channel, the free-space path loss model is used. 

\subsection{User Association Model}
Each UE associates with the communication node (i.e., BS) that provides the highest SINR. 
Let \( \delta_{u,i} \) denote the SINR of UAV-BS to UE link, and \( \delta_{\text{HIBS},i} \) is the SINR of HIBS to UE link. The association rule can be expressed as
\[
u_i^* =
\begin{cases}
\mathop{\mathrm{arg\,max}}\limits_{u \in \{1,\ldots,U\}} \delta_{u,i}, & \text{if } \delta_{u,i} \ge \delta_{\text{HIBS},i}, \\[4pt]
\text{HIBS}, & \text{otherwise.}
\end{cases}
\]
Here, \( \delta_{u,i} = \frac{P_\text{t}\|H_{u,i}\|^2}{N_0 B + I_{u,i}} \), where \(P_\text{t}\) is the transmit power, \(N_0\) is the noise power spectral density, \(B\) is the channel bandwidth, and \(I_{u,i}\) denotes the inter-UAV-BS interference term. 
Since HIBS and UAV-BSs operate in different frequency bands, \(I_{u,i}\) can be neglected, effectively reducing the expression to a signal-to-noise ratio (SNR) based association.

\subsection{User Distribution and Mobility Models} \label{sec:mobility_model}
We first categorize the total number of users $K$ into three subsets. 
The hotspot user set, denoted by $\mathcal{K}_\mathcal{H}\subset\mathcal{K}$, represent the users gathering in the predetermined emergency assembly points and remain stationary. 
The victim user set, denoted by $\mathcal{K}_\mathcal{V}\subset\mathcal{K}$, consists of users trapped under the rubble (or debris) of collapsed buildings.
The mobile user set, denoted by $\mathcal{K}_\mathcal{M}\subset\mathcal{K}$, includes the users who are primarily unaffected and mobile in the event of a disaster. 
These users are moving on the plane $z=0$ with a uniform acceleration model~\cite{mobility_model,UAV_ISAC} with an initial speed of $v_i(0)$. The heading angle and the initial velocity components are computed as $v_{x,i}(0)=v_i(0)\cos\psi_i(0), v_{y,i}(0)=v_i(0)\sin\psi_i(0)$.
Other user groups (i.e., hotspot and victim users) are assigned zero velocity.
Mobile users experience a small random acceleration to introduce trajectory randomness while maintaining overall smooth motion.
At every time slot $\Delta t$, the positions and velocities are updated as
\begin{align}
x_i(t{+}1) &= x_i(t) + v_{x,i}(t)\Delta t + \tfrac{1}{2}a_{x,i}\Delta t^2,\\ \nonumber
y_i(t{+}1) &= y_i(t) + v_{y,i}(t)\Delta t + \tfrac{1}{2}a_{y,i}\Delta t^2,\\ \nonumber
v_{x,i}(t{+}1) &= v_{x,i}(t) + a_{x,i}\Delta t,\\ \nonumber
v_{y,i}(t{+}1) &= v_{y,i}(t) + a_{y,i}\Delta t.
\end{align}
 
Additionally, the corresponding Doppler frequency is approximated as
\begin{equation}
\label{eq:rad_vel}
\mu_i(t) = \frac{2v_i(t)}{\lambda},
\end{equation}
where $v_i(t)=\sqrt{v_{x,i}^2(t)+v_{y,i}^2(t)}$.
\subsection{Transmission and Beamforming}
Each UAV-BS selects a subset of users denoted by $\mathcal{S}_u$ using SUS algorithm, such that $|\mathcal{S}_u|\leq S_{\max}$ where $S_{\max}$ is the maximum number of users that a UAV-BS can serve simultaneously.
For the selected users, the channel matrix is $H_u \in \mathbb{C}^{S_u \times N_\text{t}}$. 
Let $H_u$ be the stacked channels, then the unnormalized regularized minimum mean-square-error zero-forcing (MMSE-ZF) precoder is compute as
\begin{equation}
W_u^{\text{raw}} = H_u^H \big(H_u H_u^H + \rho I\big)^{-1},
\end{equation}
where $ H_u^H$ is Hermitian (conjugate) transpose of the channel matrix $H_u$. 
The precoder matrix $W_u^{\text{raw}}$ minimizes the mean squared error between the transmitted and received symbols while controlling noise enhancement through the regularization parameter $\rho$.  
When $\rho \!\to\! 0$, it reduces to the conventional ZF precoder.
The diagonal normalization matrix--used for minimizing inter-user interference--is given by
$D_u = \mathrm{diag}(1/\mathrm{diag}(H_u W^{\text{raw}}_u)),$
and the final beamforming matrix is computed as
\begin{equation}
W_u = \beta_u W^{\text{raw}}_u D_u,
\end{equation}
where $\beta_u$ normalizes $P_\text{t}$.
The per-user SINR becomes
\begin{equation} \label{eq:SINR}
\delta_i = 
\frac{|\mathbf{h}_{u,i}\mathbf{w}_{u,i}|^2}
{\sum_{s\neq i}|\mathbf{h}_{u,i}\mathbf{w}_{u,s}|^2+\sigma^2}.
\end{equation}

\subsection{Sensing Model}
The same communication beams are reused for monostatic radar sensing. 
For a target at range $R$ and velocity $v_r$, the time delay and Doppler shift are $\tau=\tfrac{2d}{c}$ and $\mu=\tfrac{2v_r}{\lambda}$, respectively. 
The SINR for radar processing is expressed as
\begin{align}
\delta_{\mathrm{rad}} &= 
\frac{P_\text{t}^2 G_{\mathrm{tx}}^2 \lambda^2 \sigma G_{\mathrm{proc}}}
{(4\pi)^3 R^4},
\end{align}
where $P_\text{t}$ is the UAV-BS transmit power, $\lambda$ is the carrier wavelength, $\sigma$ is the radar cross section, while $G_{\mathrm{tx}} = |a_\text{t}^{H}(\theta,\phi)w|^2$ 
represents the transmit beamforming gain in the direction $(\theta,\phi)$, and $G_{\mathrm{proc}}$ is the total processing gain. 
Furthermore, with the mobility model specified in Section~\ref{sec:mobility_model}, the speed of and distance covered by the mobile users during the entire time slots are detected.
\subsubsection{Two-UAV-BS Geometry-Based Speed Estimation} \label{kinematic-tracking}
At each $\Delta t$, the LoS direction of UAV~$u$
to the user $i$ is $\mathbf{R}_\mathcal{U}(t)=[x_i(t)-x_{u},\,y_i(t)-y_{u}]$, with the normalized vector
$\widehat{\mathbf{r}}_u(t)$.
Among all UAV-BSs, the pair $(\mathcal{U}_1,\mathcal{U}_2)$ having the largest angular separation between $\widehat{\mathbf{r}}_{\mathcal{U}_1}$ and $\widehat{\mathbf{r}}_{\mathcal{U}_2}$ is selected to ensure favorable geometric conditioning.
The radial range-rates are obtained from the inter-slot distance difference as
\begin{equation}
v_{r,u}(t)=\frac{d_{u,i}(t)-d_{u,i}(t-1)}{\Delta t}.
\end{equation}
These measurements form a linear system as
\begin{equation}
\begin{bmatrix}
\widehat{\mathbf{r}}_{\mathcal{U}_1}^\top\\[2pt]
\widehat{\mathbf{r}}_{\mathcal{U}_2}^\top
\end{bmatrix}
\mathbf{v}_i(t)=
\begin{bmatrix}
v_{r,\mathcal{U}_1}(t)\\[2pt]
v_{r,\mathcal{U}_2}(t)
\end{bmatrix},
\end{equation}
whose solution $\mathbf{v}_i(t)=[v_x(t),v_y(t)]^\top$
yields the velocity vector of the user.
The estimated total speed is
$\widehat{v}_{\mathrm{mag}}(t)=\|\mathbf{v}_i(t)\|$,
which is compared with the true speed calculated from~\eqref{eq:rad_vel}.

\subsubsection{Doppler-Based Distance Reconstruction}
For a reference UAV-BS $u_{\mathrm{ref}}$, the true range is
$d_{\mathrm{true}}(t)=d_{u_{\mathrm{ref}},i}(t)$.
The instantaneous Doppler estimate
$\mu_{\mathrm{ref}}(t)=-\dot d_{\mathrm{true}}(t)/\lambda$
is numerically integrated over time to recover the distance trajectory as
$\widehat{d}(t)
=d_{\mathrm{true}}(1)
-\lambda\!\sum_{\tau=2}^{t}\!\mu_{\mathrm{ref}}(\tau)\,\Delta t.$

\section{Methodology: Beamforming and Sensing}\label{sec:method}
This section describes in detail the methods used to implement the ISAC-over-NTN framework.
At each $\Delta t$, the locations of users, regardless of their mobility mode, are updated as described in Section~\ref{sec:mobility_model}. UAV-BSs are placed at the centers of the hotspots using a $k$-means clustering algorithm. The remaining UAV-BSs are then placed in the outer clusters, which are formed by the remaining users outside the hotspot as a result of the clustering algorithm, at appropriate positions based on the users' locations.
Following this process, after calculating the SINR of the users with respect to HIBS and UAV-BSs through~\eqref{eq:SINR}, the UAV-BSs apply MMSE-ZF beamforming and transmit multi-user beams that are jointly used for communication and sensing.
Doppler shifts due to user mobility are analyzed at each $\Delta t$, indicating the awareness of sensing within the network.
\subsection{Multi-Slot Scheduler and User Selection}

Since UAV-BSs are operating in the same spectrum, a reuse-based scheduler divides UAV-BSs into reuse groups $RG$ and transmits only one group in each mini-slot, while each one serves a maximum of $S_{\max}=2$ users simultaneously using the SUS policy, which empowers strong and mutually uncorrelated channels.
The process is summarized in Algorithm 1.
The final beamforming matrix $\mathbf{W}_u$ contains the beamforming vectors $\mathbf{w}_{u,i}$ for all scheduled users.
\begingroup
\setlength{\textfloatsep}{3pt}   
\setlength{\floatsep}{6pt}
\setlength{\intextsep}{6pt}
\captionsetup{font=small, skip=4pt} 
\begin{algorithm}
\footnotesize                    
\label{alg:multi-slot-scheduler}
\begin{algorithmic}[1]
\Require $\{H_{u,i}\}$, $U$, $K$, reuse groups $RG$, max streams $S_{\max}$, threshold $\tau_{\mathrm{SUS}}$, regularization $\rho$
\For{each outer slot $t$}
  \State UAV-BSs into $RG$ reuse groups; activate one group per mini-slot.
  \For{each active UAV-BS $u$}
    \State Sort users by power $\|\mathbf{H}_{u,i}\|^2$; set $\mathcal{S}_u\!\leftarrow\!\emptyset$.
    \For{each candidate $i$}
      \If{$|\mathbf{h}_{u,i}\mathbf{h}_{u,j}^H| < \tau_{\mathrm{SUS}}$ for all $j\!\in\!\mathcal{S}_u$} \State add $i$ to $\mathcal{S}_u$ \EndIf
      \If{$|\mathcal{S}_u| = S_{\max}$} \textbf{break} \EndIf
    \EndFor
    \State $H_u\!\leftarrow\!$ rows of selected users; $W_u^{\text{raw}}=H_u^{H}(H_u H_u^{H}+\rho I)^{-1}$
    \State $D_u=\mathrm{diag}(1/\mathrm{diag}(H_u W_u^{\text{raw}}))$; scale $W_u=\beta_u W_u^{\text{raw}} D_u$ s.t. $\mathrm{tr}(W_u W_u^{H})=P_\text{t}$
  \EndFor
  \State Compute per-user SINR $\delta_i$ and rates $R_i=B\log_2(1+\delta_i)$
\EndFor
\end{algorithmic}
\caption{SUS-Based Multi-Slot MMSE-ZF Scheduler}
\end{algorithm}
\endgroup

\subsection{Doppler--Based Motion Detection}
For each served user $i$, a normalized complex gain $g_{i,t}$ is formed at the end of every outer slot using
\begin{equation}
g_{i,t} = \frac{\mathbf{h}_{u,i}\mathbf{w}_{u,i}}{|\mathbf{h}_{u,i}\mathbf{w}_{u,i}|}.
\end{equation}
The Doppler frequency is estimated from the phase difference of the users in two consecutive slots as
\begin{equation}
\widehat{\mu}_{i,t} = 
\frac{\angle(g_{i,t}^{*}g_{i,t+1})}{2\pi\Delta t},
\end{equation}
where $\angle(g_{i,t}^{*}g_{i,t+1})$ gives the phase difference, $\phi_{i,t+1}-\phi_{i,t}$, at each time in the discrete system, spaced by $\Delta t$. 

For a single complex sinusoid in additive white Gaussian noise (AWGN) with the help of the CRB method, the variance of a frequency estimate is lower bounded as 
\begin{equation}
\mathrm{var}(\widehat{\mu}) \ge 
\frac{1}{8\pi^2\,\delta_{\mathrm{proc}}\,T_{\mathrm{obs}}^2},
\end{equation}
where $\delta_{\mathrm{proc}}$ is SINR from post-processing
and $T_{\mathrm{obs}}$ is the observation time.
With the calculation of the radial velocity from \eqref{eq:rad_vel}, and the observation interval is approximated by $\Delta t$, and the effective processing gain is absorbed into $\delta_{\mathrm{proc}}$.
Thus, a simplified variance proxy is adopted as
\begin{equation}
\sigma_{v,i}^2 \approx 
\frac{\lambda^2}{8\pi^2\,\Delta t^2\,\delta_{\mathrm{proc},i}}.
\end{equation}

The overall pipeline collaboratively integrates communication and sensing under the NTN environment. 
A reuse-based scheduler ensures fair and efficient channel access, an MMSE-ZF beamforming design provides interference-sensitive transmission, and a sense chain extracts mobility information from Doppler and CRB-weighted features. 
Thus, while MMSE-ZF beamforming provides favorable channel characteristics under imperfect channel conditions, SUS gives reliability for more stable Doppler sensing among simultaneously served users.
Together, these modules allow the ISAC-over-NTN framework that can maintain reliable connectivity even under disaster situations.

\section{Problem Formulation}\label{sec:problem}
The key factor in using the ISAC-over-NTN framework is to maximize the communication efficiency and sensing awareness of a multi-UAV-BS and HIBS network operating in a post-disaster environment, thus providing efficient communication and sensing to affected users. 
With $H_{u,i}$, and $W_u$ matrices, the system aims to represent transmission power and spatial degrees of freedom to sustain reliable coverage while providing Doppler-based mobility detection.

Let $\mathcal{B}$ the set of available terrestrial BSs.  
Each UAV-BS $u$ employs a beamforming matrix 
$\mathbf{W}_u=[\mathbf{w}_{u,1},\dots,\mathbf{w}_{u,|\mathcal{S}_u|}]\!\in\!\mathbb{C}^{N_t\times|\mathcal{S}_u|}$ 
to serve the selected user subset $\mathcal{S}_u$. 
The received signal of user $i$ is
\begin{equation}
y_i = \mathbf{h}_{u,i}\mathbf{w}_{u,i}s_i
      + \sum_{j\neq i}\mathbf{h}_{u,i}\mathbf{w}_{u,j}s_j + n_i,
\end{equation}
where $n_i\sim\mathcal{CN}(0,\sigma_n^2)$ is AWGN.  
The corresponding SINR is obtained by~\eqref{eq:SINR}, and denoted as $\delta_{u,i}$. 
The achievable downlink rate of user is $R_{u,i}=B\log_2(1+\delta_{u,i})$.

\textbf{Sensing Metric:}  
For each user $i$, the Doppler-based motion confidence 
derived from the CRB is expressed as
\begin{equation}
\mathcal{C}_i = \frac{1}{\sqrt{\sigma_{v,i}^2}}
= \frac{2\pi \Delta t}{\lambda}
\sqrt{\delta_{\mathrm{proc},i}}.
\end{equation}

\textbf{Joint Optimization Objective:}
The overall goal is to maximize the benefits of interoperable communication-sensing among all UAV-BSs and users. Thus, the joint optimization model that is developed using the weighted sum multi-objective modeling approach can be given by
\begin{align}
\max_{\{\mathbf{w}_{u,i},\,a_{u,i}\}}
\quad &
\sum_{u=1}^{U}\sum_{i=1}^{K}
\big(\eta_\text{c} R_{u,i} + \eta_\text{s} \mathcal{C}_i\big)
\label{eq:opt_main}
\\[2mm]
\text{s.t.}\quad
& \sum_{i} \|\mathbf{w}_{u,i}\|_2^2 \le P_{\text{t}}, 
&& \forall u\in\mathcal{U}, \tag{C1}\\
& \sum_{u} a_{u,i}\le 1, a_{u,i}\in\{0,1\}
&& \forall i\in\mathcal{K}, \tag{C2}\\
& |\mathbf{h}_{u,i}\mathbf{h}_{u,j}^H|^2 
   \le \tau_{\mathrm{SUS}}, 
&& \forall i\neq j. \tag{C3}
\end{align}
Here, $\eta_\text{c}$ and $\eta_\text{s}$ are weighting factors that balance communication throughput and sensing awareness.
Constraint (C1) limits the power consumed in the beam per UAV-BS, while the constraints (C2) ensure one active connection per user with the total scheduling is constant.
(C3) provides semi-orthogonality among simultaneously served users.

In the proposed framework, variables $\{a_{u,i}\}$ are determined with a maximum SNR policy, and beamforming vectors are calculated using the MMSE-ZF strategy described in Section III. The resulting suboptimal solution provides a practical balance between communication efficiency and detection awareness in realistic post-disaster conditions.

\section{Performance Evaluation}\label{sec:perf_eva}

\subsection{Performance Metrics}
For the communication performance comparison, SINR values are calculated.
To evaluate the sensing and detection performance, we set the motion-confidence score $\mathcal{C}_i = 1$. The comparisons and successes in detecting both victim and mobile users are illustrated in the confusion matrix (CM), which includes counts of true positives ($\kappa^+_\text{T}$), false negatives ($\kappa^-_\text{F}$), false positives ($\kappa^+_\text{F}$), and true negatives ($\kappa^-_\text{T}$).
For CM,  $\mathcal{A} = (\kappa^+_\text{T} + \kappa^-_\text{T})/(\kappa^+_\text{T} + \kappa^-_\text{T} + \kappa^+_\text{F}+ \kappa^-_\text{F}), \mathcal{P} = \kappa^+_\text{T}/(\kappa^+_\text{T} + \kappa^+_\text{F}), \mathcal{F} = 2(\mathcal{P}\times\mathcal{R})/(\mathcal{P}+\mathcal{R})$, where $\mathcal{A},\mathcal{P},\mathcal{R},$ and $\mathcal{F}$ are accuracy, precision, recall, and F1-score, respectively,  are calculated.
Additionally, the accuracy of Doppler-based kinematic motion tracking is evaluated by comparing the estimated and actual user velocities and distances obtained from the mobility model.

\subsection{Result and Discussion}
The environment is a 4 km$^2$ post-disaster urban area containing 10 UAV-BSs with $8\times8$ UPA MU-MIMO antenna, and a single HIBS with an altitude of $h_\text{h} =$ 20 km.
The primary consideration is the importance of NTN in communication and sensing in situations where BSs are unusable during a post-disaster event. To test this idea,
two benchmark solutions are considered in this area: first, all BSs are down, and NTN components provide network communication and sensing.
Second, NTN components are inactive, and the number of terrestrial BSs in a non-disaster environment is initially set at 40 BSs, based on the density outlined in~\cite{3GPP_Terrestrial}.
The communication performance in the environment is analyzed in scenarios where a certain ratio, denoted as $\gamma \in \{0, 0.3, 0.5, 0.8\}$, of the BSs are destroyed during the disaster. 
Additionally, the terrestrial BSs do not have sensing capabilities.
Each simulation period consists of time slots where mobile users follow a random mobility pattern, while victim and hotspot users remain stationary. 
Note that the shared hardware and spectrum in ISAC help reduce UAV-BS battery limitations without additional sensing energy costs~\cite{Dynamic_ISAC_UAV,UAV_ISAC}. Therefore, to ensure fast and reliable ISAC operation, UAV energy consumption is constrained by a fixed transmit power.
\begin{figure}
\centering
\includegraphics[width=\linewidth]{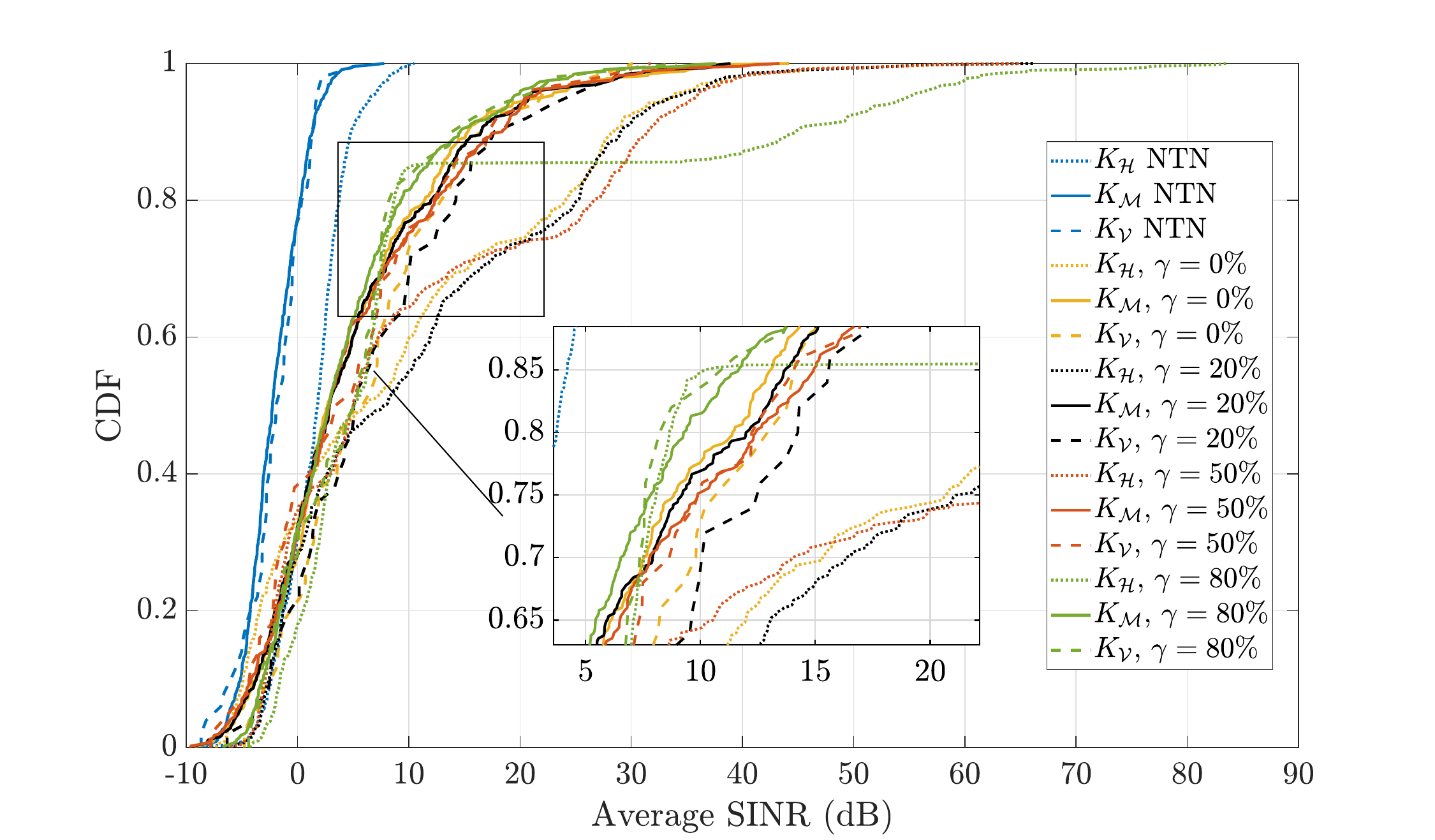}
\caption{CDF of per-user SINR for NTN and TN benchmarks under different BS failure ratios, $\gamma$.}
\label{fig:SINR}
\end{figure}

Fig.~\ref{fig:SINR} illustrates the cumulative distribution function (CDF) of the SINR for NTN scenario (all terrestrial BSs are non-functional) and for TN scenarios with different failure ratios, $\gamma$.
All three user categories are modeled for different benchmarks. 
As expected in the NTN scenario, the hotspot users (i.e., $i \in \mathcal{K}_\mathcal{H}$) achieve the highest SINR values with respect to other user types as beams are directed from UAV-BSs deployed at hotspot locations. 
However, efficient communication is also provided to the victim and mobile users (i.e., $i\in \mathcal{K}_\mathcal{V} \cup \mathcal{K}_\mathcal{M}$), for which reliable connectivity is of greater importance. 
The median SINR for all the user types is approximately -3 dB.
As a second benchmark, SINR values are varied for different $\gamma$ ratios. Here, while interference decreases as the number of BSs is reduced, the distance between the user and the BS increases, and the failure of more BSs results in a decrease in the median SINR value. This indicates that traditional TN is inadequate in the event of a real disaster, due to the failure of BSs. Furthermore, even in the lowest number of functional BS case, it is assumed that all users can be assigned to BSs without experiencing capacity problems. However, in these real-world situations, the large number of users and the failure of BSs would prevent service to a large number of users.

\begin{figure}
\centering
\includegraphics[width=.9\linewidth]{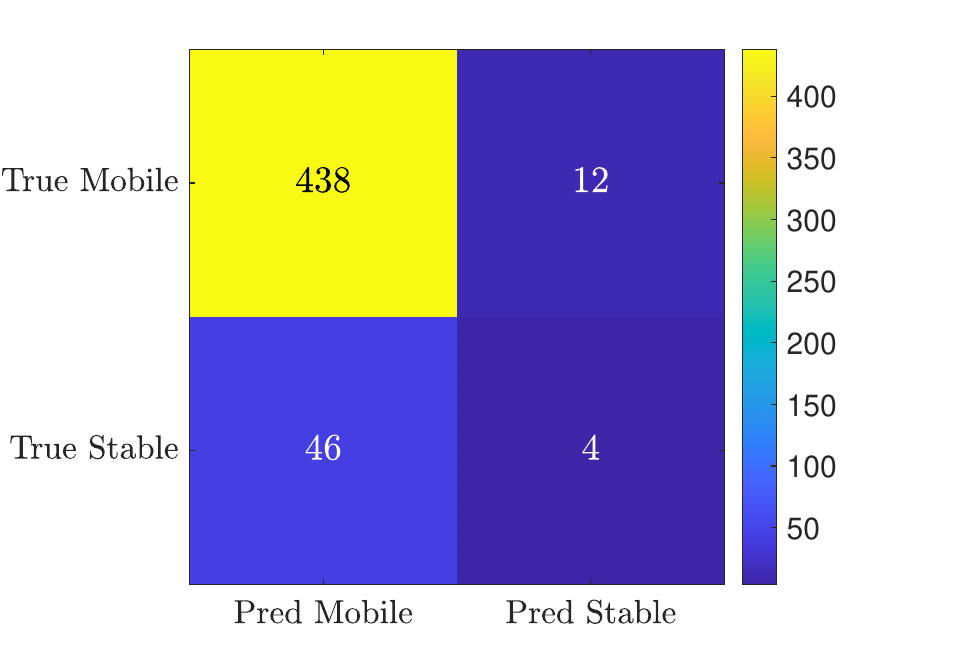}
\caption{CM for Doppler-based mobility detection with an accuracy of 0.884, precision of 0.905, and F1-score of 0.938.}
\label{fig:CM}
\end{figure}
Fig.~\ref{fig:CM} shows how successfully the victim and mobile users (i.e., $i\in \mathcal{K}_\mathcal{V} \cup \mathcal{K}_\mathcal{M}$) are detected with a Doppler motion detection design. 
In the proposed ISAC-over-NTN framework, besides providing reliable communication, detecting users who are particularly vulnerable to disasters becomes important. 
The proposed framework correctly classifies most mobile and victim users, achieving $\mathcal{P}=$ 90\%, $\mathcal{A}=$ 88\%, and $\mathcal{F}=$ 0.9 due to special beams assigned to users.

\begin{figure}
\centering
\includegraphics[width=.75\linewidth]{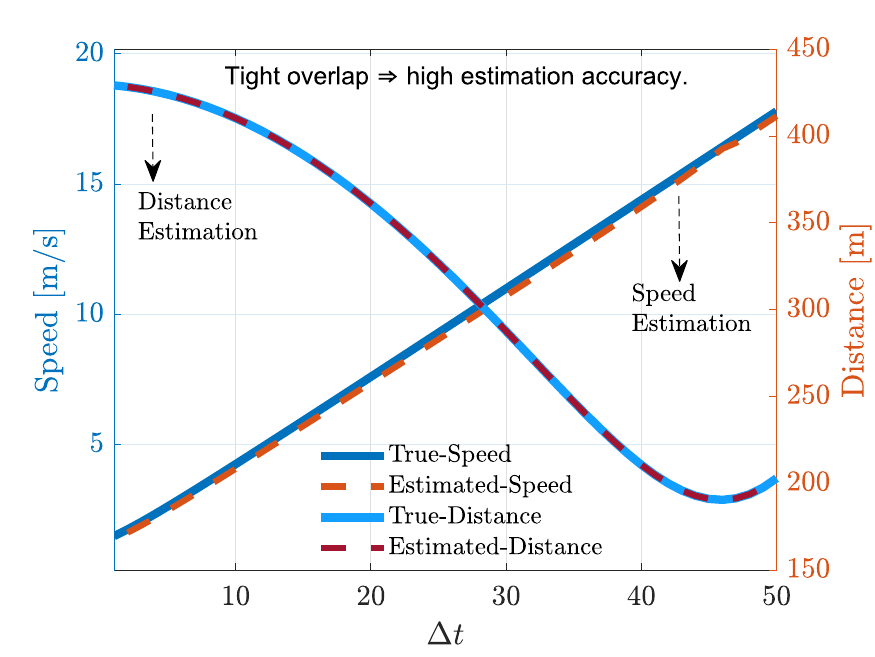}
\caption{Two-UAV-BS geometric motion tracking. (a)~Estimated versus true user speed in the left $y$-axis. (b)~True versus estimated user distance in the right $y$-axis.}
\label{fig:speed_distance}
\end{figure}
Fig. ~\ref{fig:speed_distance} presents results of kinematic tracking model, which is explained in Section ~\ref{kinematic-tracking}.
In Fig. ~\ref{fig:speed_distance}a, the speed trajectory predicted by the proposed model along the time slots matches the actual mobility trajectory generated by the mobility model.
Small deviations are either due to the nearly collinear positioning of the selected UAV-BS pair relative to the user or due to the system noise.
In Fig. ~\ref{fig:speed_distance}b, the Doppler-based distance measurement generated by the proposed model correctly follows the actual UAV-BS-user range.

These results in Figs.~\ref{fig:CM} and \ref{fig:speed_distance} together demonstrate that the proposed ISAC-over-NTN framework successfully detects users in critical situations and accurately detects both the instantaneous user speed and spatial displacement of these users, while maintaining efficient and reliable communication to users affected by the disaster in a post-disaster scenario.

\section{Conclusion}\label{sec:conc}
Disasters can cause significant damage to terrestrial infrastructure, resulting in widespread communication outages, delays in rescue operations, and difficulties in identifying victims.
To address these challenges, the ISAC-over-NTN framework proposed in this study aims to jointly ensure reliable communication and sensing awareness by utilizing an architecture comprising multiple UAV-BSs and a single HIBS.
The framework updates user locations and implements MMSE-ZF-based multi-user beamforming, while Doppler-based motion detection and reuse scheduling enable reliable communication and user mobility awareness.
Simulation results demonstrated that the proposed framework provides reliable communication and high-accuracy motion detection, achieving 90\% precision and 88\% accuracy even under BS failures. 
A natural extension of this work can be an investigation of signal-level ISAC designs with different network configurations.

\section*{Acknowledgment}
This research was supported in part by the Scientific and Technological Research Council of Türkiye (TÜBİTAK) and in part by the TÜBİTAK Informatics and Information Security Research Center (BİLGEM) under the HAPS-Com Project.

\bibliographystyle{IEEEtran}
\bibliography{main}
\end{document}